\documentclass[aps,prb,twocolumn,superscriptaddress,showpacs]{revtex4-1}
\usepackage{graphicx}
\usepackage{hyperref}
\usepackage{color}

\begin{document}
\title{Vibrations of weakly-coupled nanoparticles}

\author{Lucien Saviot}

\affiliation{Laboratoire Interdisciplinaire Carnot de Bourgogne,
UMR 5209 CNRS-Universit\'e de Bourgogne,
9 Av. A. Savary, BP 47 870, F-21078 Dijon Cedex, France}

\email{lucien.saviot@u-bourgogne.fr}

\author{Daniel B. Murray}

\affiliation{Department of Physics,
University of British Columbia Okanagan \\ 
3333 University Way, Kelowna, British Columbia, Canada V1V 1V7}

\email{daniel.murray@ubc.ca}

\begin{abstract}

The vibrations of a coupled pair of isotropic silver spheres are
investigated and compared with the vibrations of the single
isolated spheres.  Situations of both strong coupling and also
weak coupling are investigated using continuum elasticity and
perturbation theory.  The numerical calculation of the eigenmodes
of such dimers is augmented with a symmetry analysis.  This
checks the convergence and applicability of the numerical method
and shows how the eigenmodes of the dimer are constructed from
those of the isolated spheres.  The frequencies of the lowest
frequency vibrations of such dimers are shown to be very
sensitive to the strength of the coupling between the spheres.
Some of these modes can be detected by inelastic light scattering
and time-resolved optical measurements which provides a convenient
way to study the nature of the mechanical coupling in dimers of
micro and nanoparticles.
\end{abstract}

\maketitle

\section{Introduction}

The ensemble or acoustic vibrations of micro and nanoparticles
have been studied during the last few decades using a variety of
spectroscopic and time-resolved optical techniques.  These
vibrations can be used for the characterization of the size and
shape of the particles.  Their coupling with the electrons are
known to contribute to the excitonic dephasing\cite{Takagahara96}
and even to have a universal role in the optical emission of
quantum dots.\cite{OronPRL09} Recently some experimental
measurements have focused on the acoustic modes of a pair of
mechanically coupled metallic nanoparticles
(NP)\cite{LiNL05,TchebotarevaCPC09}
or silica microspheres.\cite{DehouxOL09} Measurements also exist
for other systems where the mechanical coupling is more complex
such as nanocolumns made of overlapping spheres\cite{BurginNL08},
self-assembled systems for which the coherent vibrations of the
lattice of spherical NPs have been evoked\cite{Courty2005,IvandaAST08}
and nanopowders where individual NPs are in
contact.\cite{PighiniJNR06,MurrayJNO06,SaviotPRB08} The goal of
the present theoretical study of the vibrations of dimers of NPs
is to improve the understanding of experimental results
concerning dumbbell NPs and also to pave the way to studies of a
variety of other more complex systems where individual NPs are
close enough so that their vibrations can be coupled.

A free homogeneous isotropic sphere has vibrational modes which
we denote using our notation from a previous work.\cite{SaviotPRB09}
Briefly, these modes are classified according to four integer
quantum numbers.  The first is the distinction between torsional
(T, zero divergence) and spheroidal (S, nonzero divergence) modes.
Modes are next labeled by the usual nonnegative integer angular
momentum $\ell$.  There is also an angular momentum $z$-component
$m$.  Finally, modes are indexed by integer $n$ in order of
increasing frequency, starting with 1.  We thus denote an
arbitrary normal mode of a sphere either as T$_{\ell,m}^n$ or
S$_{\ell,m}^n$.

For a silver nanosphere whose diameter is small compared to the wavelength
of light, S$_0$ and S$_2$ are the only Raman active vibrations and S$_0$
are the only vibrations observed by time-resolved pump-probe experiments.
Since the fundamental vibrations are the main features in both kind of
experiments, this paper will mainly focus on the case of symmetric
vibrations originating from the fundamental S$_{0,0}$ and S$_{2,0}$.

\section{Method}

\subsection{Continuum eigenvibrations and symmetry}

Hathorn \textit{et al.}\cite{HathornJPCA02} used a molecular
dynamics model to investigate the vibrations of polymer
nanoparticle dimers.  In the present work, we use a continuum
elastic model which has been shown to be suitable for
microspheres down to rather small nanospheres.\cite{CombePRB09}
Moreover, it enables a straightforward way to identify the
symmetry of the eigenvibrations (irreducible representations) and
to compare with the vibrations of free spheres.  We use the tools
recently presented elsewhere\cite{SaviotPRB09} which are suitable
to calculate the eigenvibrations of arbitrary systems and to
classify them in terms of symmetry, volume variation and
projections. The volume integrations required in the calculation
of the eigenmodes according to the method introduced by Visscher
\textit{et al.}\cite{visscher} were computed numerically by
integration only along the $z$ direction by taking advantage of
the axial symmetry.  The wavefunctions were expanded in a
$x^i y^j z^k$ basis with $i + j + k \le N$.  We used $N=20$ in
this work unless stated otherwise.

The main focus of this work is the case of dimers consisting of
weakly coupled spheres.  The eigenmodes of any system satisfy two
basic rules: they belong to an irreducible representation of the
point group of interest and all the points of the system must
oscillate at the same frequency for a given eigenmode. These two
rules let us predict the nature of the eigenmodes in the weak
coupling regime.

The point group associated with a dimer NP is
D$_{\infty h}$ if it is made of two identical spheres and
C$_{\infty v}$ if the spheres are different. The irreducible
representations are A$_{1}$, A$_{2}$, E$_{1}$,
E$_{2}$, E$_{3}$, \ldots for C$_{\infty v}$ and the
same with parity (u and g) for D$_{\infty h}$.
The degeneracy with $m$ for systems having spherical symmetry is
partially lifted for axisymmetric systems according to the following rule:
S$_{\ell,0}^n$ modes turn into A$_{1}$,
T$_{\ell,0}^n$ modes turn into A$_{2}$ and
S$_{\ell,m \ne 0}^n$ and T$_{\ell,m \ne 0}^n$ modes turn into
E$_{|m|}$.
These rules are obtained by checking the character table of
C$_{\infty v}$ and remembering the $e^{im\phi}$ dependence of the
displacements.

The three rigid translations and three rigid rotations of the
individual spheres can be considered to be vibrational modes
with zero frequency.  Conventionally, they are not included
when enumerating modes.  For our purposes here, it is very
helpful to include them.  We label them as S$_{1,m}^0$ and
T$_{1,m}^0$ respectively.  This is an exception to our convention
that modes are indexed with $n$ starting from 1.  They transform
using the rules given before for an axisymmetric system.

\subsection{Perturbation theory}

\subsubsection{Basic equations}

Consider two elastic objects, where the normal modes of
vibration are known for each of them individually.  
In this section, we show how a linear perturbation expansion
can be used to find the frequency shift of modes
when two such objects are weakly coupled.  There are several
specific situations we have in mind.  The first is where
the two objects are two nanoparticles which are weakly
coupled together.  The second is where a nanoparticle is
weakly coupled to a substrate.  However, this same formalism
could be applied to a system of three or more objects.  For
clarity, we will always refer to a system of two objects
below, even though other numbers of objects are possible

The total number of atoms in the system of two objects is
$N$.  The index $i$ labels the atoms from 1 to $N$.
The Cartesian axes are labeled with Greek indices
$\alpha$ or $\beta$, going from 1 to 3.  $u_{i \alpha}$ is
the displacement of atom $i$ along the $\alpha$ axis.
$m_i$ is the mass of atom $i$.

Without coupling between the objects, the net force on atom $i$
along axis $\alpha$ is 
\begin{equation}
\label{e405}
\sum_{j \beta} B_{i \alpha j \beta} u_{j \beta}
\end{equation}
where the dynamical matrix of the uncoupled system is
$B_{i \alpha j \beta}$.  When the two objects are coupled, the
dynamical matrix changes to $A_{i \alpha j \beta}$, where
$\lambda C_{i \alpha j \beta}$ is the dynamical matrix due to
the coupling alone and $\lambda$ is a scalar parameter that we
can use for a perturbative expansion. 
\begin{equation}
\label{e410}
A_{i \alpha j \beta} = B_{i \alpha j \beta} + \lambda C_{i \alpha j \beta}
\end{equation}

Consider a normal mode of the uncoupled system.  It has frequency
$\omega_o$ and atomic displacements $u_{o i \alpha}$.  The
dynamical equation is 
\begin{equation}
\label{e420}
\sum_{j \beta} B_{i \alpha j \beta} u_{o j \beta} = -\omega_o^2 m_i u_{o i \alpha}
\end{equation}

When $\lambda$ is made nonzero, the mode will continuously shift
to new atomic displacements $u_{i \alpha}$ and new frequency
$\omega$.  The new dynamical equation is 
\begin{equation}
\label{e430}
\sum_{j \beta} A_{i \alpha j \beta} u_{j \beta} = -\omega^2 m_i u_{i \alpha}
\end{equation}

We expand the square of the mode frequency as a power series in
$\lambda$ as follows: 
\begin{equation}
\label{e440}
\omega^2 = \omega_o^2 + \lambda \omega_1^2 + \lambda^2 \omega_2^2 + ...
\end{equation}
and likewise expand the mode displacements in $\lambda$:
\begin{equation}
\label{e450}
u_{i \alpha} = u_{o i \alpha} + \lambda u_{1 i \alpha}+ \lambda^2 u_{2 i \alpha} + ...
\end{equation}

We keep the mode displacements normalized as follows: 
\begin{equation}
\label{e460}
\sum_{i \alpha} m_i u_{o i \alpha} u_{o i \alpha} = \sum_{i \alpha} m_i u_{i \alpha} u_{i \alpha} 
\end{equation}

To determine the effect of the coupling term, we
substitute the expansions in $\lambda$ into
Eq.~(\ref{e430}).  Collecting all terms linear in $\lambda$,
we obtain 
\begin{eqnarray}
\label{e470}
\nonumber
\sum_{j \beta} C_{i \alpha j \beta} u_{o j \beta} + \sum_{j \beta} B_{i \alpha j \beta} u_{1 j \beta} \\
= -m_i ( \omega_1^2 u_{o i \alpha} + \omega_0^2 u_{1 i \alpha} )
\end{eqnarray}

Equation~(\ref{e470}) is now multiplied through by
$u_{o i \alpha}$ and each term is summed over $i$ and $\alpha$.
In addition, substitution of the expansions in $\lambda$
into Eq. (\ref{e460}) and collecting linear terms in $\lambda$
tells us that 
\begin{equation}
\label{e475}
\sum_{i \alpha} m_i u_{o i \alpha} u_{1 i \alpha} = 0
\end{equation}

Consequently, we obtain an expression for the mode frequency
to linear order in $\lambda$ and set $\lambda$ to one. 
\begin{equation}
\label{e480}
\omega^2 \simeq \omega_o^2 - 
\frac{\sum_{i \alpha j \beta}C_{i \alpha j \beta} u_{o i \alpha} u_{o j \beta}}{\sum_{i \alpha}m_i u_{o i \alpha}^2 }
\end{equation}

\subsubsection{Two-point coupling}

We now specialize to the case where the two objects are coupled
in the simplest possible way.  We want to connect them by a
``spring''.  What we mean by ``spring'' will be explained below.
In order for the coupling to be weak, we restrict its influence to a very
small volume fraction of both objects as in
Ref.~\onlinecite{TchebotarevaCPC09}. We idealize this situation to the
case of a coupling involving just a
single atom on each object.  Atom $a$ on the first object is
coupled through the spring to atom $b$ on the second
object.  Such an arrangement is only capable of
coupling the component of the force which is parallel
to the axis of the spring.  We will suppose that the
spring is aligned along the $z$-axis.  Thus, atoms
$a$ and $b$ share the same $x$ and $y$ coordinates.

In this situation, there are only four nonzero
elements of the coupling matrix:
$C_{a3a3}$, $C_{a3b3}$, $C_{b3a3}$, and $C_{b3b3}$.
Let $F_{a3}$ denote the $z$-component of the force
on atom $a$ from the spring.  Then
\begin{equation}
F_{a3} = C_{a3a3} u_{a3} + C_{a3b3} u_{b3}  
\end{equation}

and
\begin{equation}
F_{b3} = C_{b3a3} u_{a3} + C_{b3b3} u_{b3}  
\end{equation}

The first order perturbation formula for the frequency
now becomes

\begin{eqnarray}
\nonumber
\omega^2 \simeq \omega_o^2 - 
\frac{ u_{oa3}^2 C_{a3a3} + u_{ob3}^2 C_{b3b3} + 2 u_{oa3} u_{ob3} C_{a3b3}}{ \sum_{i \alpha} m_i u_{o i \alpha}^2 } \\
\end{eqnarray}

\subsubsection{Thin cylinder}

We now consider the case of a ``spring'' consisting of a thin
cylinder.  If the material has Young's modulus $Y$ then the
spring constant of this spring is $k_{sp} = Y A / L$ where $A$ is
the cross sectional area of the cylinder and $L$ is the length.
However, this needs to be modified if the frequency of the system is not low
compared to the internal vibrational modes of the thin cylinder because the
cylinder does not behave as an ideal massless spring in that case.
In this section, we analyse the situation when the frequency is
not necessarily low.

Atom $a$ is located at the top end of the cylinder, at $z = L/2$.
Atom $b$ is located at $z= -L/2$.  $u_z(z,t)$ is the displacement
field inside the cylinder.  It has the general form
\begin{equation}
u_z(z,t) = B e^{i(kz-\omega t)} + D e^{i(-kz-\omega t)} 
\end{equation}
where $B$ and $D$ are constants, $k$ = $\omega/v$ and
$v = \sqrt{Y/\rho}$ is the speed of longitudinal
vibrations in a thin rod, 2872~m/s in silver.

$e_{zz}(z,t)$ is the strain in the cylinder, given by
$e_{zz} = \partial u_z / \partial z$.
The stress is $\sigma_{zz}(z,t)$, given by
$\sigma_{zz} = Y e_{zz}$.
\begin{eqnarray}
F_{a3} e^{-i \omega t} = -A \sigma_{zz}(L/2,t)   \\
F_{b3} e^{-i \omega t} = A \sigma_{zz}(-L/2,t)   \\
u_{a3} e^{-i \omega t} = u_z(L/2,t)              \\
u_{b3} e^{-i \omega t} = u_z(-L/2,t)
\end{eqnarray}

For simplicity, we restrict the remainder of our discussion to
the case where $u_{a3} = -u_{b3}$ and $F_{a3} = -F_{b3}$ which is
valid for the symmetric modes we are mainly interested in.
Note that $C_{a3a3}-C_{a3b3}$ = $F_{a3}/u_{a3}$. Furthermore, $B+D=0$ and
\begin{equation}
u_{a3} = B( e^{i k L/2} - e^{-i k L/2} )
\end{equation}
\begin{equation}
F_{a3} = -i k Y A B (  e^{i k L/2} + e^{-i k L/2} )
\end{equation}
\begin{equation}
C_{a3a3}-C_{a3b3} =
\frac{-i k Y A (  e^{i k L/2} + e^{-i k L/2} )}{ e^{i k L/2} - e^{-i k L/2} }
\end{equation}
\begin{equation}
C_{a3a3}-C_{a3b3} = - k Y A / \tan(k L / 2)
\end{equation}

The first order perturbation formula is
\begin{equation}
\omega^2 \simeq \omega_o^2 +
\frac{ 2 u_{oa3}^2 k Y A }{ \sum_{i \alpha} m_i u_{o i \alpha}^2 \tan(k L / 2)}
\end{equation}

Finally, we take the continuum limit and apply this to
the case where the two objects are two identical
homogeneous spheres.  Each sphere has mass
$M_{sph}$ and volume $V_{sph}$.  In this case,
\begin{equation}
\label{perturb}
\omega^2 \simeq \omega_o^2 +
\frac{p k Y A}{M_{sph} \tan(k L / 2)}
\end{equation}
where we define $p$ as
\begin{equation}
\label{e720}
p = \frac{(u_z(\vec{r}_a))^2 V_{sph}}{\int_{sph} \vec{u} \cdot \vec{u} dV}
\end{equation}

Here are some values of $p$ for the north pole of an isotropic
silver sphere.  For the zero frequency translation mode along $z$, $p=1$.
For the spheroidal mode with $m=0$ and $\ell$ equal to 0, 2, 3 and 4, $p$
equals 0.87, 3.26, 6.07 and 9.18 respectively for the silver
spheres considered in this work.

\section{Results and Discussion}

\subsection{Symmetrical dimer}

We consider a dimer made of slightly overlapping identical
spheres of radius $R=5$~nm whose centers are on the $z$ axis at
$z = d/2$ and $z = -d/2$.
The perturbation approach presented before does not deal with such a
coupling between nanoparticles but its simplicity makes it a better
starting point to understand how the vibrations of a dimer are built.
In all this work, we chose to work
only with an isotropic approximation for silver (mass density:
10.5~g/cm$^3$, sound speeds: 3747~m/s (longitudinal) and 1740~m/s
(transverse)). Such an approximation is known to be adequate in most cases
and in particular in the case of multiply twinned particles. Therefore
elastic anisotropy which has been shown only recently to play a significant
role for mono-domain gold nanoparticles\cite{PortalesPNAS08} and never for
silver ones will be ignored here.
Table~\ref{TabSphere} gives the frequencies and
irreducible representations of the lowest oscillations of a
single sphere.  Table~\ref{TabSym} gives the assignments for the
lowest frequency modes of a dimer made of the same spheres having
their center being 9~nm apart ($d < 2 R$).

\begin{table*}
\begin{tabular}{c|ccccccccccccccc}
$\nu$ (GHz)& 138.5&  146.9  &  202.2  &  214.0  &
 219.4  &  281.7  &  282.2  &  286.4  &  319.2  &
 335.1  &  340.1  &  347.1  &  375.8  &  395.2  &
 396.5\\
\hline
i. r.   & T$_2^1$ & S$_2^1$ & S$_1^1$ & T$_3^1$ &
S$_3^1$ & S$_4^1$ & T$_4^1$ & S$_2^2$ & T$_1^1$ &
S$_0^1$ & S$_5^1$ & T$_5^1$ & S$_3^2$ & T$_2^2$ &
S$_6^1$\\
\end{tabular}
\caption{\label{TabSphere}Frequencies of the lowest frequency modes of
an isotropic silver sphere having a radius of 5~nm calculated using the
model by Lamb.}
\end{table*}

\begin{table}
\begin{center}
\begin{tabular}{c|c|c|c|c}
$i$ & $\nu$ (GHz) & i.r. & decomposition & $\Delta\nu$ (\%)\\
\hline
     7 &  32.9  & A$_{2u}$ & T$_{1,0}^0$     $\ominus$ T$_{1,0}^0$     & 2.1\\
   8-9 &  37.8  & E$_{1u}$ & T$_{1,\pm 1}^0$ $\ominus$ T$_{1,\pm 1}^0$ & 2.2\\
    10 &  67.4  & A$_{1g}$ & S$_{1,0}^0$     $\ominus$ S$_{1,0}^0$     & 1.0\\
 11-12 &  78.1  & E$_{1g}$ & T$_{1,\pm 1}^0$ $\oplus$  T$_{1,\pm 1}^0$ & 0.8\\
    13 & 138.9  & A$_{2g}$ & T$_{2,0}^1$     $\ominus$ T$_{2,0}^1$     & 0.0\\
 14-15 & 140.3  & E$_{2g}$ & T$_{2,\pm 2}^1$ $\ominus$ T$_{2,\pm 2}^1$ & 0.2\\
 16-17 & 143.6  & E$_{1u}$ & T$_{2,\pm 1}^1$ $\ominus$ T$_{2,\pm 1}^1$ & 0.1\\
 18-19 & 144.2  & E$_{2u}$ & T$_{2,\pm 2}^1$ $\oplus$  T$_{2,\pm 2}^1$ & 0.1\\
 20-21 & 145.2  & E$_{1g}$ & S$_{2,\pm 1}^1$ $\oplus$  S$_{2,\pm 1}^1$ & 0.0\\
 22-23 & 147.3  & E$_{2g}$ & S$_{2,\pm 2}^1$ $\oplus$  S$_{2,\pm 2}^1$ & 0.0 \\
    24 & 148.6  & A$_{1u}$ & S$_{2,0}^1$     $\ominus$ S$_{2,0}^1$     & 0.0 \\
 25-26 & 149.8  & E$_{2u}$ & S$_{2,\pm 2}^1$ $\ominus$ S$_{2,\pm 2}^1$ & 0.1 \\
    27 & 154.0  & A$_{2u}$ & T$_{2,0}^1$     $\oplus$  T$_{2,0}^1$     & 0.3 \\
 28-29 & 156.8  & E$_{1u}$ & S$_{2,\pm 1}^1$ $\ominus$ S$_{2,\pm 1}^1$ & 0.2 \\
 30-31 & 168.9  & E$_{1g}$ & T$_{2,\pm 1}^1$ $\oplus$  T$_{2,\pm 1}^1$ & 0.1 \\
    32 & 181.5  & A$_{1g}$ & S$_{2,0}^1$     $\oplus$  S$_{2,0}^1$     & 0.2 \\
\ldots & \ldots & \ldots          & \ldots                     &\ldots   \\
   112 & 314.5  & A$_{1g}$ & S$_{4,0}^1$     $\oplus$  S$_{4,0}^1$     & 0.1 \\
\ldots & \ldots & \ldots          & \ldots                      &\ldots  \\
   119 & 336.7  & A$_{1u}$ & S$_{0,0}^1$     $\ominus$ S$_{0,0}^1$     & 0.0 \\
\ldots & \ldots & \ldots          & \ldots                     &\ldots   \\
   134 & 343.8  & A$_{1g}$ & S$_{0,0}^1$     $\oplus$  S$_{0,0}^1$     & 0.1 \\
\ldots & \ldots & \ldots          & \ldots                   &\ldots     \\
   141 & 347.5  & A$_{1u}$ & S$_{5,0}^1$     $\oplus$ S$_{5,0}^1$      & 0.0 
\end{tabular}
\end{center}
\caption{\label{TabSym} Modes of a symmetric dimer made of two
overlapping isotropic silver spheres whose radius is 5~nm where
the distance between the sphere centers is 9~nm. In-phase and
out-of-phase oscillations of the spheres are labeled with
$\oplus$ and $\ominus$ respectively. The last column shows the
decrease of the calculated frequencies when increasing $N$ from
18 to 20 so as to illustrate the numerical convergence.
}
\end{table}

In the weak coupling regime, the eigenmodes of a dimer can be
seen as the superposition of one eigenmode for each sphere. Weak
coupling means that the vibration of each sphere is almost
unaffected by the presence of the other. For a symmetrical dimer,
due to the inversion symmetry, the eigenmodes of the dimer have
to be either even or odd which means that the vibrations of both
spheres have to be identical and either in-phase (identical
displacements after translating one sphere over the other) or
out-of-phase (opposite displacements).

\begin{figure}
\includegraphics[width=0.45\columnwidth]{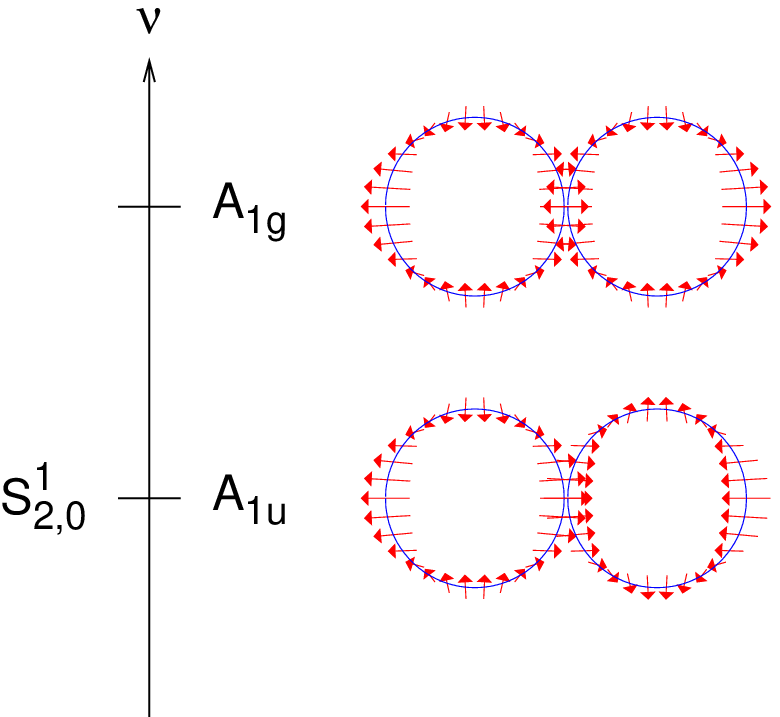}\hfill
\includegraphics[width=0.45\columnwidth]{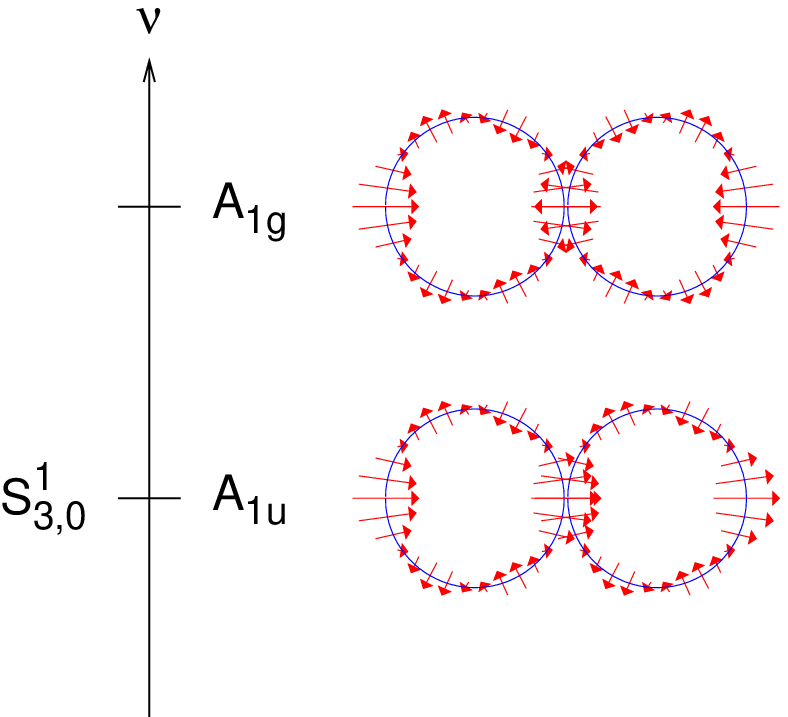}
\caption{\label{A1g}Odd and even vibrations of the dimer constructed
from S$_{\ell,0}^1$ modes of the free spheres with $\ell=2$ (left) and
$\ell=3$ (right). The $z$-axis goes through the centers of both spheres
and the 3D displacements are obtained by rotation around this axis.}
\end{figure}

As a first example, we focus on the quadrupolar mode
S$_{2,0}^1$ (A$_{1g}$ for D$_{\infty h}$). Two eigenmodes
of the dimer are obtained from this mode: one having the
A$_{1g}$ symmetry for which the spheres oscillate in-phase
($i=32$ in Table~\ref{TabSym}) and an A$_{1u}$ ($i=24$) one for
out-of-phase oscillations. These modes are illustrated in Fig.~\ref{A1g}.
Due to the coupling, the oscillation is shifted in frequency with
respect to that of the free spheres. If the vibrations of the two
spheres result in different displacements for the center of the
dimer, then the frequency will be increased. Conversely, if the
two displacements are identical, then there will be almost no
frequency shift because the displacement of each sphere is
unaffected by the other one. By decreasing the coupling between
the two spheres (\textit{i.e.} increasing the distance between
their centers) the frequency difference between the two
modes of the dimer discussed above is reduced as they both evolve
towards the frequency of the S$_{2,0}^1$ mode of the free
spheres.  It should be noted that close to $d = 2 R$, the various
frequencies do not reach those for a free sphere. This can be
attributed to the singular nature of this point together with the
limitation of this continuum elasticity model which is
unrealistic for very small overlapping volumes.

Because the parity of the spheroidal vibrations is the same as
the parity of $\ell$, the construction of the odd and even
vibrations of the dimer is different for odd $\ell$ as
illustrated in Fig.~\ref{A1g} for $\ell=3$ but the resulting even
vibration of the dimer is always the one having different
displacements for the two spheres at the center of the dimer and
therefore the one having a larger frequency shift compared to the
frequency of the free spheres.

\begin{figure}
\includegraphics[width=\columnwidth]{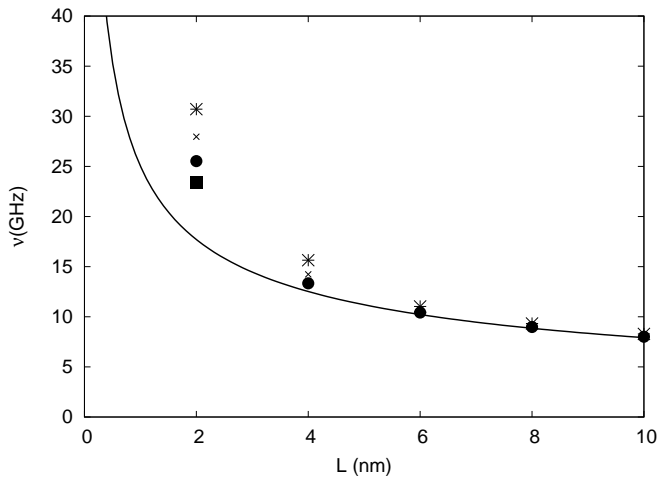}
\caption{\label{spring}Frequency of the lowest A$_{1g}$ mode
calculated for two silver spheres of radii $R=5$~nm connected
with a cylinder of radius $r = R/10 = 0.5$~nm and length
$L$ ($N=20$ full circles, $N=18$ crosses, $N=16$ stars and $N=21$
full squares) compared to the frequency of two point masses
connected with a spring having a force constant equal to that of
the cylinder (line) using Eq.~\ref{perturb}.}
\end{figure}

The translation along $z$ for a single sphere corresponds to a
vibration at zero frequency. From this mode, we can construct an
in-phase oscillation of the dimer which corresponds to the
translation along $z$ of the dimer (0~GHz). The out-of-phase
oscillation corresponds to the lowest A$_{1g}$ mode (67.4~GHz).
To confirm this assignment we checked that the frequency of this
A$_{1g}$ mode tends to zero as the coupling between the spheres
decreases. In order to have a finer control the magnitude of this
coupling, we considered a dumbbell made of spheres connected by a
cylinder whose radius is $r=R/10$. This small dimension ensures
that the eigenfrequencies of the free cylinder are large which
restricts the number of vibrations of the cylinder which can
manifest in the low frequency range.  Figure~\ref{spring} shows
the variation of the frequency of the lowest frequency A$_{1g}$
mode as a function of the length of the cylinder $L$.  Since the
spheres are hardly deformed but rather translated during the
oscillation, it is possible to model this oscillation as that of
two point masses connected by a spring. While a first order
perturbation theory for a massless spring is possible, we present
here only a basic derivation valid for this particular vibration.
The force constant of the massless equivalent spring equivalent
to the cylinder is $k_{sp} = Y \frac{A}{L}$ where $Y$ is the
Young's modulus and the cross sectional area of the cylinder is
$A = \pi r^2$. This expression is valid only for $L \gg r$.
The resulting frequency for the dimer is then
$\omega = \sqrt{2 k_{sp} / M}$.  The very good agreement observed
in Fig.~\ref{spring} for $L>4$~nm is an indication of both the
validity of the assignment of this mode and the accuracy of the
computation of the vibration eigenmodes even for such a
challenging geometry. This particular oscillation is of interest
since it was recently experimentally observed.\cite{TchebotarevaCPC09}
The present work demonstrates that measuring its frequency is
equivalent to determining the force constant of the ``spring''
which connects the two NPs constituting the dimer.

Table~\ref{TabSym} presents a set of low-frequency modes $i=1-12$
originating from the translations (S$_{1,m}^0$) and rotations
(T$_{1,m}^0$) of the spheres. Modes $i=1-6$ (not shown)
correspond to the rotations and translations of the dimer with
zero frequency.  These 12 eigenmodes correspond to the number of
rotations and translations of the two spheres.  Then modes
$i=13-32$ correspond to combinations of the S$_{2,m}^1$ or
T$_{2,m}^1$ and so on. Again, these 20 eigenmodes correspond to
the number of S$_{2,m}^1$ and T$_{2,m}^1$ eigenmodes of the two
spheres.  Modes corresponding to the superpositions of the
breathing modes (S$_{0,0}^1$) are also shown. However, since the
vibrational density of states increases with frequency, even in
this relatively low coupling case this mode mixes significantly
with (S$_{5,0}^1$).  This mixing is not apparent in
Table~\ref{TabSym} since only the main contributions are shown.
The same rule regarding the shift of the frequencies as discussed
for the modes originating from S$_{2,0}^1$ before applies for
every modes.

The way the vibrations of a dimer are built is the analog of the
way molecular orbitals are built using the linear combination of
atomic orbitals (LCAO) method. Such an analogy has already been
pointed out before in the case of acoustic waves in periodic
ensemble of spheres in a host material.\cite{Kafesaki95} For
example, the A$_{1u}$ and A$_{1g}$ modes built from the breathing
mode of the spheres (S$_{0,0}^1$) are analogous to the bonding
and anti-bonding molecular orbital of H$_2$ built from the $1s$
orbitals. However the reduction in the energy of the bonding
orbital compared to that of the $1s$ atomic orbital does not
exist for the A$_{1u}$ vibration since there is no acoustic
equivalence for the screening of the charges of the nuclei due to
the electrons.

\begin{figure} \includegraphics[width=\columnwidth]{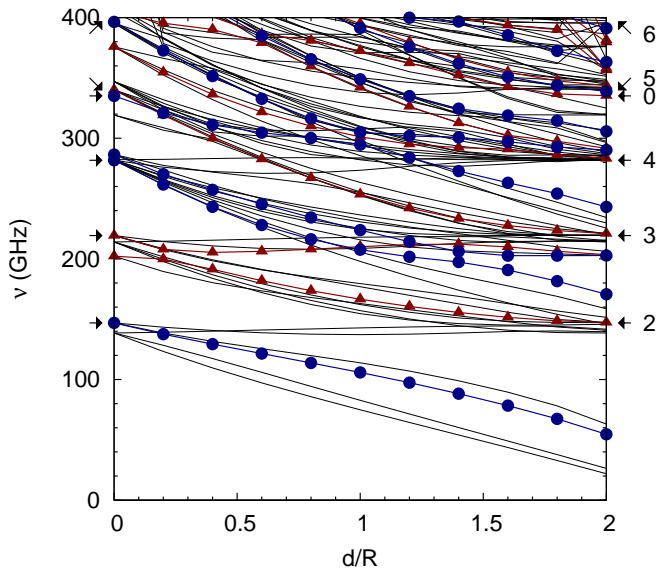}
\caption{\label{coupling}Variation of the frequencies of a dumbbell made of
two silver spheres with radii $R=5$~nm with varying distance $d$ between
their centers. The A$_{1g}$ and A$_{1u}$ branches are plotted with circles
(blue online) and triangles (red online) respectively. The frequencies of
some S$^1_\ell$ modes are marked with arrows and labeled with $\ell$.}
\end{figure}

In order to study the influence of the coupling of the vibrations of the
two spheres, the variation of the frequencies of the different vibrations
are plotted in Fig.~\ref{coupling} as a function of the distance between
the center of the spheres.  For $d=0$ the system consists of a single
sphere and the frequencies match those reported in Table~\ref{TabSphere}.
For $d$ very close to $2 R$, the behavior is singular because for $d=2R$
the two spheres touch at a single point. Moreover,  for $d=2R-\epsilon$
each sphere feels the other sphere as a mass attached to it but it is no
longer the case for $d=2R+\epsilon$. So despite the the convergence being
good, the interpretation of the results is not obvious at this point since
we can expect a non-smooth variation of the frequencies at $d=2R$.
Figure~\ref{coupling} presents the branches corresponding to the situation
when $d < 2R$. The lines between the calculated frequencies connect
eigenmodes having the same irreducible representation.  This was also done
in a previous work.\cite{SaviotPRB09} The mixings between the branches
having the same irreducible representations are numerous and show that the
nature of the vibrations in the case of a strong coupling can be quite
complex.

\subsection{Convergence of the numerical method}

\begin{figure}
\includegraphics[width=\columnwidth]{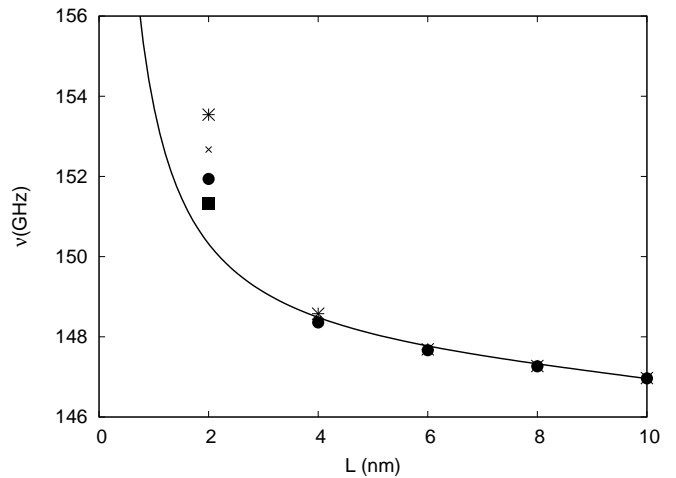}
\caption{\label{massspring}Frequency of the second lowest
A$_{1g}$ mode calculated for two silver spheres of radii $R=5$~nm
connected with a narrow cylinder of radius $r = R/10 = 0.5$~nm
and length $L$ ($N=20$ full circles, $N=18$ crosses, $N=16$ stars
and $N=21$ full squares) compared to the first order perturbation
calculation of the coupling of the S$_{2,0}^1$ eigenmodes of the
spheres through the extensional eigenmode of the thin cylinder
(line) using Eq.~\ref{perturb}.}
\end{figure}

The numerical method used in this work has been shown to be very
accurate for several different geometries. The question of its
reliability for the challenging systems in this work is addressed
here.  To that end, we performed the same calculations with
$N=18$ (Table~\ref{TabSym}) and $N=16$ and $N=21$
(Figs.~\ref{spring} and \ref{massspring}) to compare with the
reference $N=20$. This comparison does not rely on the mode index
$i$ but rather on the irreducible representation. For example,
$\Delta \nu$ for mode $i=10$ in Table~\ref{TabSym} is the
frequency variation for the lowest frequency A$_{1g}$ mode with
$N=20$ and $N=18$. We varied $N$ by steps of 2 because adding or
removing one to $N$ does not change the frequencies of all the
eigenmodes.  This is due to the inversion symmetry. Changing $N$
by one adds or removes even or odd functions only depending on
the parity of $N$.  Therefore only the convergence of even or odd
modes is changed.

\begin{figure}
\includegraphics[width=\columnwidth]{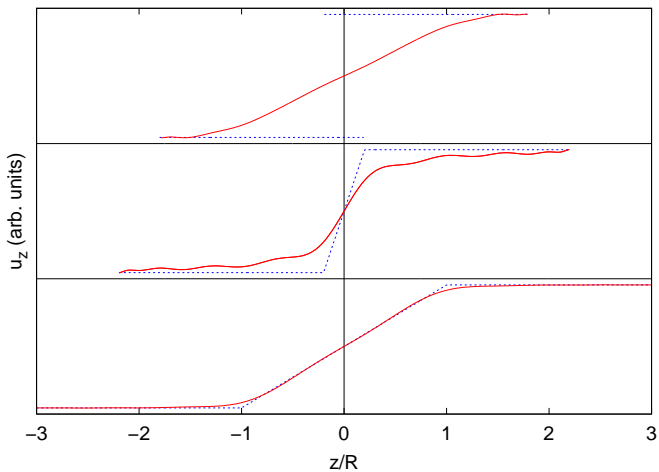}
\caption{\label{spconv}Normalized displacements along the $z$-axis for
the lowest frequency A$_{1g}$ mode of a dimer of overlapping spheres
with d/R=1.6 (top) and of a cylindrical-necked dumbbell with cylinder
length $L$ equal to 2~nm (middle) and 10~nm (bottom). Each sphere has
radius 5~nm.
The continuous lines correspond to the numerical results for $N=20$
while the dotted ones correspond to a constant displacement inside
each sphere (translation along $z$) and a linear variation of the
displacement inside the cylinder when it is present.}
\end{figure}

The convergence for the results presented in Table~\ref{TabSym}
is good.  This is due to the use of a rather high value for $N$.
The largest variations $\Delta \nu$ are obtained for the lowest
frequency modes.  This tendency is confirmed by
Figs.~\ref{spring}, \ref{massspring} and \ref{spconv} except
for large values of $L$ for which the convergence for the lowest
A$_{1g}$ modes is much better. The worst convergence is for
$L=2$~nm for which the variation between $N=20$ and $N=21$ in
Fig.~\ref{spconv} is still quite large (9\%).  This can be
understood when comparing the calculated displacements along the
$z$-axis with that expected from the ideal spring model for which both
spheres are translated in opposite directions and the displacement varies
linearly inside the spring (see
Fig.~\ref{spconv}). The steep variation of this displacement
inside the cylinder together with the flat variation inside the
spheres are much more difficult to reproduce with power functions
of $z$ when the length of
the spring ($L$) is small. Moreover, for $L=2$~nm, $L/r = 4$ so the
behavior of the cylinder may differ from the spring approximation.  The
variation of the displacement for overlapping spheres is not so steep which
explains why the convergence is much better in that case.  Indeed, the
spring approximation does not hold as can be checked from the displacement
plotted in Fig.~\ref{spconv} (top). The deformation of each sphere is
quite significant or in other words the coupling is not so weak.  Similar
comments apply for the other modes which are combinations of rotations and
translations of the spheres with a non-zero frequency except that the
relevant characteristic value of the cylinder is not related to its
stretching but rather to torsion and bending. For the other modes at higher
frequency, the displacements inside the spheres are not constant which can
be easier to reproduce with power functions of $x$, $y$ and $z$. Therefore,
the convergence can be better than for the eigenmodes made of the rotations
and translations of the two spheres.

The convergence for the second lowest A$_{1g}$ vibration was also checked
using the same method. In that case, the calculations are compared with the
first order perturbation results presented before for a thin cylinder. This
is necessary because the extensional vibration of this cylinder has a
frequency similar to the S$_2^1$ vibration of the spheres for $L=2R=10$~nm.
The comparison of both calculations in Fig.~\ref{massspring} shows the
very high accuracy of the RUS calculations for long enough cylinders,
\textit{i.e.} when the approximations made for the perturbation theory are
valid.

\subsection{Non-symmetrical dimer}

In this general case, there are some restrictions on which isolated sphere
eigenvibrations can be combined to create a vibration eigenmode for the
dimer. First, the two spheres' vibrations should share the same irreducible
representation in C$_{\infty v}$ (same $m$ and same character (S or T) for
$m=0$) otherwise their superposition would not have a well-defined
symmetry. Second, their frequencies should match because all the points
inside the system oscillate at the same frequency for an eigenmode.  

When going from a symmetric dimer to a non-symmetric one, the inversion
symmetry is lost and the two spheres have different sets of vibrational
frequencies. For example it is possible that the frequency of the
S$^1_{1,0}$ mode of one sphere might happen to match the frequency of the
S$^1_{2,0}$ mode of the other. In this case, some modes of the dimer as a
whole would be hybridizations of these two. However, such an exact match
of mode frequencies is very unlikely. Leaving aside such coincidences, in
the very weak coupling regime the modes of the dimer are simply the
superposition of one eigenmode for one sphere with the other sphere at rest
because that is the only way to have all the points of the dimer
oscillating at the same frequency. When the coupling becomes stronger the
modes can be mixed even more than in the case of the symmetric dimer since
parity is lost. In that case, any modes having the same value of $m$ can
mix provided their frequencies are close except for $m=0$ for which
torsional and spheroidal modes can not mix.

\section{Conclusion}

We have demonstrated that an exhaustive description of the vibrations of a
weakly coupled dimer based on continuum calculations, perturbation theory
and symmetry considerations is possible.  While the strong coupling regime
can be investigated using these numerical tools, it is of course quite hard
to interpret the results since the free oscillations of the spheres
constituting the dimer are strongly modified. The convergence of the
calculations is discussed in the weak coupling regime thanks to the
comparison with perturbation calculations and accurate results are obtained
in most cases. General equations were given to extend the perturbation
approach to other weakly coupled systems. This work paves the way for the
understanding of the vibrations of more complex systems of interest such as
chains of overlapping spheres as in nanocolumns\cite{BurginNL08} and chains
of non-spherical particles.

\bibliography{dimer}

%Merlin.mbs v4.21 2009-07-09.
\begin{thebibliography}{10}%
\makeatletter
\providecommand \@ifxundefined [1]{%
 \ifx #1\undefined \expandafter \@firstoftwo
 \else \expandafter \@secondoftwo
\fi
}%
\providecommand \@ifnum [1]{%
 \ifnum #1\expandafter \@firstoftwo
 \else \expandafter \@secondoftwo
\fi
}%
\providecommand \enquote [1]{``#1''}%
\providecommand \bibnamefont  [1]{#1}%
\providecommand \bibfnamefont [1]{#1}%
\providecommand \citenamefont [1]{#1}%
\providecommand\href[0]{\@sanitize\@href}%
\providecommand\@href[1]{\endgroup\@@startlink{#1}\endgroup\@@href}%
\providecommand\@@href[1]{#1\@@endlink}%
\providecommand \@sanitize [0]{\begingroup\catcode`\&12\catcode`\#12\relax}%
\@ifxundefined \pdfoutput {\@firstoftwo}{%
 \@ifnum{\z@=\pdfoutput}{\@firstoftwo}{\@secondoftwo}%
}{%
 \providecommand\@@startlink[1]{\leavevmode\special{html:<a href="#1">}}%
 \providecommand\@@endlink[0]{\special{html:</a>}}%
}{%
 \providecommand\@@startlink[1]{%
  \leavevmode
  \pdfstartlink
   attr{/Border[0 0 1 ]/H/I/C[0 1 1]}%
   user{/Subtype/Link/A<</Type/Action/S/URI/URI(#1)>>}%
  \relax
 }%
 \providecommand\@@endlink[0]{\pdfendlink}%
}%
\providecommand \url  [0]{\begingroup\@sanitize \@url }%
\providecommand \@url [1]{\endgroup\@href {#1}{\urlprefix}}%
\providecommand \urlprefix [0]{URL }%
\providecommand \Eprint[0]{\href }%
\@ifxundefined \urlstyle {%
  \providecommand \doi [1]{doi:\discretionary{}{}{}#1}%
}{%
  \providecommand \doi [0]{doi:\discretionary{}{}{}\begingroup
  \urlstyle{rm}\Url }%
}%
\providecommand \doibase [0]{http://dx.doi.org/}%
\providecommand \Doi[1]{\href{\doibase#1}}%
\providecommand \bibAnnote [3]{%
  \BibitemShut{#1}%
  \begin{quotation}\noindent
    \textsc{Key:}\ #2\\\textsc{Annotation:}\ #3%
  \end{quotation}%
}%
\providecommand \bibAnnoteFile [2]{%
  \IfFileExists{#2}{\bibAnnote {#1} {#2} {\input{#2}}}{}%
}%
\providecommand \typeout [0]{\immediate \write \m@ne }%
\providecommand \selectlanguage [0]{\@gobble}%
\providecommand \bibinfo [0]{\@secondoftwo}%
\providecommand \bibfield [0]{\@secondoftwo}%
\providecommand \translation [1]{[#1]}%
\providecommand \BibitemOpen[0]{}%
\providecommand \bibitemStop [0]{}%
\providecommand \bibitemNoStop [0]{.\EOS\space}%
\providecommand \EOS [0]{\spacefactor3000\relax}%
\providecommand \BibitemShut [1]{\csname bibitem#1\endcsname}%
%</preamble>
\bibitem{Takagahara96}%
  \BibitemOpen
  \bibfield{author}{%
  \bibinfo {author} {\bibfnamefont{T.}~\bibnamefont{Takagahara}},\ }%
  \bibfield{journal}{%
  \bibinfo {journal} {Phys. Rev. Lett.}\ }%
  \textbf{\bibinfo {volume} {71}},\ \bibinfo {pages} {3577} (\bibinfo {year}
  {1993})%
  \bibAnnoteFile{NoStop}{Takagahara96}%
\bibitem{OronPRL09}%
  \BibitemOpen
  \bibfield{author}{%
  \bibinfo {author} {\bibfnamefont{D.}~\bibnamefont{Oron}}, \bibinfo {author}
  {\bibfnamefont{A.}~\bibnamefont{Aharoni}}, \bibinfo {author}
  {\bibfnamefont{C.}~\bibnamefont{de~Mello~Donega}}, \bibinfo {author}
  {\bibfnamefont{J.}~\bibnamefont{van Rijssel}}, \bibinfo {author}
  {\bibfnamefont{A.}~\bibnamefont{Meijerink}},\ and\ \bibinfo {author}
  {\bibfnamefont{U.}~\bibnamefont{Banin}},\ }%
  \bibfield{journal}{%
  \bibinfo {journal} {Phys. Rev. Lett.}\ }%
  \textbf{\bibinfo {volume} {102}},\ \bibinfo {pages} {177402} (\bibinfo {year}
  {2009})%
  \bibAnnoteFile{NoStop}{OronPRL09}%
\bibitem{LiNL05}%
  \BibitemOpen
  \bibfield{author}{%
  \bibinfo {author} {\bibfnamefont{Y.}~\bibnamefont{Li}}, \bibinfo {author}
  {\bibfnamefont{Q.}~\bibnamefont{Zhang}}, \bibinfo {author}
  {\bibfnamefont{A.~V.}\ \bibnamefont{Nurmikko}},\ and\ \bibinfo {author}
  {\bibfnamefont{S.}~\bibnamefont{Sun}},\ }%
  \bibfield{journal}{%
  \bibinfo {journal} {Nano Lett.}\ }%
  \textbf{\bibinfo {volume} {5}},\ \bibinfo {pages} {1689} (\bibinfo {year}
  {2005})%
  \bibAnnoteFile{NoStop}{LiNL05}%
\bibitem{TchebotarevaCPC09}%
  \BibitemOpen
  \bibfield{author}{%
  \bibinfo {author} {\bibfnamefont{A.~L.}\ \bibnamefont{Tchebotareva}},
  \bibinfo {author} {\bibfnamefont{M.~A.}\ \bibnamefont{{van Dijk}}}, \bibinfo
  {author} {\bibfnamefont{P.~V.}\ \bibnamefont{Ruijgrok}}, \bibinfo {author}
  {\bibfnamefont{V.}~\bibnamefont{Fokkema}}, \bibinfo {author}
  {\bibfnamefont{M.~H.~S.}\ \bibnamefont{Hesselberth}}, \bibinfo {author}
  {\bibfnamefont{M.}~\bibnamefont{Lippitz}},\ and\ \bibinfo {author}
  {\bibfnamefont{M.}~\bibnamefont{Orrit}},\ }%
  \bibfield{journal}{%
  \bibinfo {journal} {Chem. Phys. Chem.}\ }%
  \textbf{\bibinfo {volume} {10}},\ \bibinfo {pages} {111} (\bibinfo {year}
  {2009})%
  \bibAnnoteFile{NoStop}{TchebotarevaCPC09}%
\bibitem{DehouxOL09}%
  \BibitemOpen
  \bibfield{author}{%
  \bibinfo {author} {\bibfnamefont{T.}~\bibnamefont{Dehoux}}, \bibinfo {author}
  {\bibfnamefont{T.~A.}\ \bibnamefont{Kelf}}, \bibinfo {author}
  {\bibfnamefont{M.}~\bibnamefont{Tomoda}}, \bibinfo {author}
  {\bibfnamefont{O.}~\bibnamefont{Matsuda}}, \bibinfo {author}
  {\bibfnamefont{O.~B.}\ \bibnamefont{Wright}}, \bibinfo {author}
  {\bibfnamefont{K.}~\bibnamefont{Ueno}}, \bibinfo {author}
  {\bibfnamefont{Y.}~\bibnamefont{Nishijima}}, \bibinfo {author}
  {\bibfnamefont{S.}~\bibnamefont{Juodkazis}}, \bibinfo {author}
  {\bibfnamefont{H.}~\bibnamefont{Misawa}}, \bibinfo {author}
  {\bibfnamefont{V.}~\bibnamefont{Tournat}},\ and\ \bibinfo {author}
  {\bibfnamefont{V.~E.}\ \bibnamefont{Gusev}},\ }%
  \bibfield{journal}{%
  \bibinfo {journal} {Opt. Lett.}\ }%
  \textbf{\bibinfo {volume} {34}},\ \bibinfo {pages} {3740} (\bibinfo {year}
  {2009})%
  \bibAnnoteFile{NoStop}{DehouxOL09}%
\bibitem{BurginNL08}%
  \BibitemOpen
  \bibfield{author}{%
  \bibinfo {author} {\bibfnamefont{J.}~\bibnamefont{Burgin}}, \bibinfo {author}
  {\bibfnamefont{P.}~\bibnamefont{Langot}}, \bibinfo {author}
  {\bibfnamefont{A.}~\bibnamefont{Arbouet}}, \bibinfo {author}
  {\bibfnamefont{J.}~\bibnamefont{Margueritat}}, \bibinfo {author}
  {\bibfnamefont{J.}~\bibnamefont{Gonzalo}}, \bibinfo {author}
  {\bibfnamefont{C.~N.}\ \bibnamefont{Afonso}}, \bibinfo {author}
  {\bibfnamefont{F.}~\bibnamefont{Vall{\'e}e}}, \bibinfo {author}
  {\bibfnamefont{A.}~\bibnamefont{Mlayah}}, \bibinfo {author}
  {\bibfnamefont{M.~D.}\ \bibnamefont{Rossell}},\ and\ \bibinfo {author}
  {\bibfnamefont{G.}~\bibnamefont{{Van Tendeloo}}},\ }%
  \bibfield{journal}{%
  \bibinfo {journal} {Nano Lett.}\ }%
  \textbf{\bibinfo {volume} {8}},\ \bibinfo {pages} {1296} (\bibinfo {year}
  {2008})%
  \bibAnnoteFile{NoStop}{BurginNL08}%
\bibitem{Courty2005}%
  \BibitemOpen
  \bibfield{author}{%
  \bibinfo {author} {\bibfnamefont{A.}~\bibnamefont{Courty}}, \bibinfo {author}
  {\bibfnamefont{A.}~\bibnamefont{Mermet}}, \bibinfo {author}
  {\bibfnamefont{P.~A.}\ \bibnamefont{Albouy}}, \bibinfo {author}
  {\bibfnamefont{E.}~\bibnamefont{Duval}},\ and\ \bibinfo {author}
  {\bibfnamefont{M.~P.}\ \bibnamefont{Pileni}},\ }%
  \bibfield{journal}{%
  \bibinfo {journal} {Nature Materials}\ }%
  \textbf{\bibinfo {volume} {4}},\ \bibinfo {pages} {395} (\bibinfo {year}
  {2005})%
  \bibAnnoteFile{NoStop}{Courty2005}%
\bibitem{IvandaAST08}%
  \BibitemOpen
  \bibfield{author}{%
  \bibinfo {author} {\bibfnamefont{M.}~\bibnamefont{Ivanda}}, \bibinfo {author}
  {\bibfnamefont{M.}~\bibnamefont{Buljan}}, \bibinfo {author}
  {\bibfnamefont{U.~V.}\ \bibnamefont{Desnica}}, \bibinfo {author}
  {\bibfnamefont{K.}~\bibnamefont{Furi\'c}}, \bibinfo {author}
  {\bibfnamefont{D.}~\bibnamefont{Risti\'c}}, \bibinfo {author}
  {\bibfnamefont{G.~C.}\ \bibnamefont{Righini}},\ and\ \bibinfo {author}
  {\bibfnamefont{M.}~\bibnamefont{Ferrari}},\ }%
  \bibfield{journal}{%
  \bibinfo {journal} {Advances in Science and Technology}\ }%
  \textbf{\bibinfo {volume} {55}},\ \bibinfo {pages} {127} (\bibinfo {year}
  {2008})%
  \bibAnnoteFile{NoStop}{IvandaAST08}%
\bibitem{PighiniJNR06}%
  \BibitemOpen
  \bibfield{author}{%
  \bibinfo {author} {\bibfnamefont{C.}~\bibnamefont{Pighini}}, \bibinfo
  {author} {\bibfnamefont{D.}~\bibnamefont{Aymes}}, \bibinfo {author}
  {\bibfnamefont{N.}~\bibnamefont{Millot}},\ and\ \bibinfo {author}
  {\bibfnamefont{L.}~\bibnamefont{Saviot}},\ }%
  \bibfield{journal}{%
  \bibinfo {journal} {J. Nanoparticle Research}\ }%
  \textbf{\bibinfo {volume} {9}},\ \bibinfo {pages} {309} (\bibinfo {year}
  {2007})%
  \bibAnnoteFile{NoStop}{PighiniJNR06}%
\bibitem{MurrayJNO06}%
  \BibitemOpen
  \bibfield{author}{%
  \bibinfo {author} {\bibfnamefont{D.~B.}\ \bibnamefont{Murray}}, \bibinfo
  {author} {\bibfnamefont{C.~H.}\ \bibnamefont{Netting}}, \bibinfo {author}
  {\bibfnamefont{L.}~\bibnamefont{Saviot}}, \bibinfo {author}
  {\bibfnamefont{C.}~\bibnamefont{Pighini}}, \bibinfo {author}
  {\bibfnamefont{N.}~\bibnamefont{Millot}}, \bibinfo {author}
  {\bibfnamefont{D.}~\bibnamefont{Aymes}},\ and\ \bibinfo {author}
  {\bibfnamefont{H.-L.}\ \bibnamefont{Liu}},\ }%
  \bibfield{journal}{%
  \bibinfo {journal} {J. Nanoelectron. Optoelectron.}\ }%
  \textbf{\bibinfo {volume} {1}},\ \bibinfo {pages} {92} (\bibinfo {year}
  {2006})%
  \bibAnnoteFile{NoStop}{MurrayJNO06}%
\bibitem{SaviotPRB08}%
  \BibitemOpen
  \bibfield{author}{%
  \bibinfo {author} {\bibfnamefont{L.}~\bibnamefont{Saviot}}, \bibinfo {author}
  {\bibfnamefont{C.~H.}\ \bibnamefont{Netting}}, \bibinfo {author}
  {\bibfnamefont{D.~B.}\ \bibnamefont{Murray}}, \bibinfo {author}
  {\bibfnamefont{S.}~\bibnamefont{Rols}}, \bibinfo {author}
  {\bibfnamefont{A.}~\bibnamefont{Mermet}}, \bibinfo {author}
  {\bibfnamefont{A.-L.}\ \bibnamefont{Papa}}, \bibinfo {author}
  {\bibfnamefont{C.}~\bibnamefont{Pighini}}, \bibinfo {author}
  {\bibfnamefont{D.}~\bibnamefont{Aymes}},\ and\ \bibinfo {author}
  {\bibfnamefont{N.}~\bibnamefont{Millot}},\ }%
  \bibfield{journal}{%
  \bibinfo {journal} {Phys. Rev. B}\ }%
  \textbf{\bibinfo {volume} {78}},\ \bibinfo {pages} {245426} (\bibinfo {year}
  {2008})%
  \bibAnnoteFile{NoStop}{SaviotPRB08}%
\bibitem{SaviotPRB09}%
  \BibitemOpen
  \bibfield{author}{%
  \bibinfo {author} {\bibfnamefont{L.}~\bibnamefont{Saviot}}\ and\ \bibinfo
  {author} {\bibfnamefont{D.~B.}\ \bibnamefont{Murray}},\ }%
  \bibfield{journal}{%
  \bibinfo {journal} {Phys. Rev. B}\ }%
  \textbf{\bibinfo {volume} {79}},\ \bibinfo {pages} {214101} (\bibinfo {year}
  {2009})%
  \bibAnnoteFile{NoStop}{SaviotPRB09}%
\bibitem{HathornJPCA02}%
  \BibitemOpen
  \bibfield{author}{%
  \bibinfo {author} {\bibfnamefont{B.~C.}\ \bibnamefont{Hathorn}}, \bibinfo
  {author} {\bibfnamefont{B.~G.}\ \bibnamefont{Sumpter}}, \bibinfo {author}
  {\bibfnamefont{D.~W.}\ \bibnamefont{Noid}}, \bibinfo {author}
  {\bibfnamefont{R.~E.}\ \bibnamefont{Tuzun}},\ and\ \bibinfo {author}
  {\bibfnamefont{C.}~\bibnamefont{Yang}},\ }%
  \bibfield{journal}{%
  \bibinfo {journal} {J. Phys. Chem. A}\ }%
  \textbf{\bibinfo {volume} {106}},\ \bibinfo {pages} {9174} (\bibinfo {year}
  {2002})%
  \bibAnnoteFile{NoStop}{HathornJPCA02}%
\bibitem{CombePRB09}%
  \BibitemOpen
  \bibfield{author}{%
  \bibinfo {author} {\bibfnamefont{N.}~\bibnamefont{Combe}}, \bibinfo {author}
  {\bibfnamefont{P.-M.}\ \bibnamefont{Chassaing}},\ and\ \bibinfo {author}
  {\bibfnamefont{F.}~\bibnamefont{Demangeot}},\ }%
  \bibfield{journal}{%
  \bibinfo {journal} {Phys. Rev. B}\ }%
  \textbf{\bibinfo {volume} {79}},\ \bibinfo {pages} {045408} (\bibinfo {year}
  {2009})%
  \bibAnnoteFile{NoStop}{CombePRB09}%
\bibitem{visscher}%
  \BibitemOpen
  \bibfield{author}{%
  \bibinfo {author} {\bibfnamefont{W.~M.}\ \bibnamefont{Visscher}}, \bibinfo
  {author} {\bibfnamefont{A.}~\bibnamefont{Migliori}}, \bibinfo {author}
  {\bibfnamefont{T.~M.}\ \bibnamefont{Bell}},\ and\ \bibinfo {author}
  {\bibfnamefont{R.~A.}\ \bibnamefont{Reinert}},\ }%
  \bibfield{journal}{%
  \bibinfo {journal} {J. Acoust. Soc. Am.}\ }%
  \textbf{\bibinfo {volume} {90}},\ \bibinfo {pages} {2154} (\bibinfo {year}
  {1991})%
  \bibAnnoteFile{NoStop}{visscher}%
\bibitem{PortalesPNAS08}%
  \BibitemOpen
  \bibfield{author}{%
  \bibinfo {author} {\bibfnamefont{H.}~\bibnamefont{Portal\`es}}, \bibinfo
  {author} {\bibfnamefont{N.}~\bibnamefont{Goubet}}, \bibinfo {author}
  {\bibfnamefont{L.}~\bibnamefont{Saviot}}, \bibinfo {author}
  {\bibfnamefont{S.}~\bibnamefont{Adichtchev}}, \bibinfo {author}
  {\bibfnamefont{D.~B.}\ \bibnamefont{Murray}}, \bibinfo {author}
  {\bibfnamefont{A.}~\bibnamefont{Mermet}}, \bibinfo {author}
  {\bibfnamefont{E.}~\bibnamefont{Duval}},\ and\ \bibinfo {author}
  {\bibfnamefont{M.-P.}\ \bibnamefont{Pil\'eni}},\ }%
  \bibfield{journal}{%
  \bibinfo {journal} {Proc. Natl. Acad. Sci. U.S.A}\ }%
  \textbf{\bibinfo {volume} {105}},\ \bibinfo {pages} {14784} (\bibinfo {year}
  {2008})%
  \bibAnnoteFile{NoStop}{PortalesPNAS08}%
\bibitem{Kafesaki95}%
  \BibitemOpen
  \bibfield{author}{%
  \bibinfo {author} {\bibfnamefont{M.}~\bibnamefont{Kafesaki}}\ and\ \bibinfo
  {author} {\bibfnamefont{E.~N.}\ \bibnamefont{Economou}},\ }%
  \bibfield{journal}{%
  \bibinfo {journal} {Phys. Rev. B}\ }%
  \textbf{\bibinfo {volume} {52}},\ \bibinfo {pages} {13317} (\bibinfo {year}
  {1995})%
  \bibAnnoteFile{NoStop}{Kafesaki95}%
\end{thebibliography}%

\end{document}